# THE EXCHANGE PARAMETERS IN LANTHANUM MANGANITES

B.V. Karpenko[*], A.V. Kuznetzov[**]

[*]Institute of physics of metals, Ural department of Russian Academy of Sciences, Ekaterinburg 620041, Russia
[**]Ural state university, Ekaterinburg 620083, Russia

Address for correspondence: Karpenko Boris Victorovich, Institute of physics of metals, URO RAN, Ekaterinburg 620041, phone 378-35-64, FACS (343)3745244, e-mail:naumov@imp.uran.ru

The spin-wave spectrum in the metallic ferromagnetic lanthanum cubic manganites was investigated. The interactions with four coordination spheres (32 neighbors) were taken into account. The Heisenberg superexchange between all manganese ions as well as double exchange between the manganese ions with different valence were under consideration. The obtained magnon spectrum permits to investigate dispersion relations along an any crystallographic directions. This possibility was used for the numerical evaluation of the exchange parameters in the compounds $La_{0.75}Ca_{0.25}MnO_3$ and $La_{0.70}Ca_{0.30}MnO_3$ from the inelastic neutron scattering data.

Исследован спин-волновой спектр металлических ферромагнитных манганитов, обладающих кубической структурой, с учетом взаимодействий вплоть до четвертой координационной сферы (32 соседа). Приняты во внимание гайзенберговские сверхобменные взаимодействия между всеми ионами марганца и двойной обмен между разновалентными ионами. Полученный магнонный спектр позволяет исследовать дисперсионные соотношения вдоль любых кристаллографических направлений. Эта возможность использована для оценки численных значений обменных параметров в соединениях $La_{0.75}Ca_{0.25}MnO_3$ и $La_{0.70}Ca_{0.30}MnO_3$ путем обработки дисперсионных кривых, полученных с помощью неупругого нейтронного рассеяния.

Среди различных методов изучения манганитов с переменной валентностью важное место занимают исследования элементарных магнитных возбуждений с помощью нейтронного рассеяния и его теоретической интерпретации [1-14]. В недавних работах [12,13] магнонный спектр в металлических ферромагнитных соединениях $La_{0.75}Ca_{0.25}MnO_3$ (LCMO25) и $La_{0.70}Ca_{0.30}MnO_3$ (LCMO30) был исследован методом неупругого нейтронного рассеяния в главных кристаллографических направлениях. Результаты экспериментов представлены на рисунках 1 и 2. Авторы этих работ, анализируя данные с помощью метода обобщенной восприимчивости, используя гамильтониан гайзенберговского вида, вводя усредненный спин для разновалентных ионов марганца, учитывая взаимодействия лишь с первой и четвертой координационными сферами, получили численные значения обменных интегралов для ближайших соседей $I_1$ и соседей четвертого порядка $I_4$ в кубическом кристалле при бесщелевых дисперсионных кривых. Результаты их расчетов также показаны на рисунках 1 и 2.

Поскольку исследуемые соединения являются металлами и, следовательно, в них реализуется механизм двойного обмена, имеющий негайзенберговский характер, то, на наш взгляд, гайзенберговское приближение, принятое в работах [12,13], по крайней мере, проблематично. Поэтому мы решили вновь обратиться к экспериментальным данным работ [12,13] и исследовать их с использованием моделей двойного обмена (негайзенберговское взаимодействие) и сверхобмена (гайзенберговское взаимодействие) одновременно. При этом мы будем учитывать взаимодействия со всеми четырьмя координационными сферами (а не только с первой и четвертой как в работах [12,13]). Мы не будем также вводить понятие усредненного спина, в отличие от [12,13]. При анализе дисперсионных кривых мы будем пользоваться формулами для магнонного спектра, полученными нами ранее в работе [14].

Итак, энергия спиновой волны дается выражением [14]:

$$E(\vec{q}) = 2\{J - \sum_{p=1}^{4} \gamma_p(\vec{q}) J_p + A\}, \qquad (1)$$

где $\vec{q}$ есть квазиимпульс спиновой волны, индекс $p$ нумерует координационные сферы и где

$$\gamma_p(\vec{q}) = \sum_{j=1}^{z_p} \exp(-i\vec{q}\,^j\vec{\Delta}_p), \qquad (2)$$

$z_p$ - число узлов в $p$-ой сфере, $^j\vec{\Delta}_p$ - радиус-вектор $j$-го узла в $p$-ой координационной сфере. Другие обозначения:

$$J_p = (1-x)^2\left(S+\frac{1}{2}\right)(^{11}I_p) + x^2 S(^{22}I_p) + x(1-x)\sqrt{2S(2S+1)}\left[^{12}I_p + \frac{B_p}{(2S+1)^2}\right], \qquad (3)$$

где $x$ - концентрация четырехвалентных ионов марганца, обладающих спином $S$ ($S=3/2$); $^{11}I_p, ^{22}I_p, ^{12}I_p$ - обменные интегралы с соседом $p$-го порядка соответственно для пар $Mn^{3+}$-$Mn^{3+}$, $Mn^{4+}$-$Mn^{4+}$ и $Mn^{3+}$-$Mn^{4+}$; $B_p$ - интеграл переноса к соседу $p$-го порядка между ионами $Mn^{3+}$ и $Mn^{4+}$. Далее:

$$J = \sum_{p=1}^{4} z_p \tilde{J}_p, \qquad (4)$$

$$\tilde{J}_p = J_p + f(S)\left[^{12}I + \frac{B_p}{(2S+1)^2}\right], \quad f(S) = 2S + \frac{1}{2} - \sqrt{2S(2S+1)}, \qquad (5)$$

$$A = (1-x)\left(S+\frac{1}{2}\right)K_1 + xSK_2, \qquad (6)$$

где $K_1$ и $K_2$ - константы одноионной анизотропии соответственно для трех и четырех валентных ионов марганца. Ниже мы будем считать $A=0$ в соответствии с данными из работ [12,13].

Далее, для определения численных значений параметров взаимодействия $J_p; p=1,2,3,4$ мы будем пользоваться методом наименьших квадратов, используя экспериментальные данные (кружки) из рисунков 1 и 2. В качестве варьируемых параметров у нас будут, естественно, четыре величины $J_p$, а также величина $J$ (см. формулы (1) и (4)). С использованием направлений (100)+(110)+(111) получаем в единицах мэв для LCMO25

$$J_1 = 4.402, \; J_2 = -0.097, \; J_3 = -0.068, \; J_4 = 0.3, \; J = 26.44 \qquad (7)$$

и для LCMO30

$$J_1 = 3.635,\ J_2 = -0.087,\ J_3 = -0.081,\ J_4 = 0.586,\ J = 23.755 \qquad (8)$$

Расчетные кривые с параметрами (7) и (8) приведены на рисунках 1 и 2. Если для обработки использовать все шесть направлений из рисунков 1 и 2, то получаются значения близкие к (7) и (8).

Сравним (7) и (8) с результатами работы [13]. Там получено:

$$I_1 = 2.088,\ I_2 = 0,\ I_3 = 0,\ I_4 = 0.136 \qquad (9)$$

для LCMO25 и

$$I_1 = 1.719,\ I_2 = 0,\ I_3 = 0,\ I_4 = 0.335 \qquad (10)$$

для LCMO30.

Нулевые значения для $I_2$ и $I_3$ в (9) и (10) были приняты авторами [13] априори. Наши $J_p$ являются аналогами величин $I_p$ из [13], но буквальное их сравнение невозможно: $J_p$ имеют "внутреннюю структуру" (см. формулу(3)), тогда как $I_p$ представляют собой некоторые усредненные параметры при усредненных спинах, равных 1.875 для LCMO25 и 1,85 для LCMO30. Величины $J_1$ и $J_4$ примерно вдвое превосходят $I_1$ и $I_4$ главным образом из-за того, что в выражения для $J_1$ и $J_4$ входит величина спина $S$.

Из формул (7) и (8) видно, что интегралы $J_2$ и $J_3$ действительно малы по абсолютной величине по сравнению с $J_4$, однако, поскольку число соседей второго порядка равно 12, а третьего порядка равно 8 по сравнению с 6-ю соседями четвертого порядка, то суммарный вклад второго и третьего порядка сравним с величиной четвертого порядка и им не следует пренебрегать.

Отметим одну трудность, возникающую при численных оценках обменных параметров. Суммарное ферромагнитное взаимодействие, мерой которого может служить величина $J$, определяемая формулой (4), оказалась большей для LCMO25, чем для LCMO30 (см. (7) и (8)). Это находится в противоречии с тем фактом, что температура Кюри $T_c$ у LCMO25 ниже, чем у LCMO30. То же самое можно сказать и о результатах работы [13]: $I_1 + I_4$ и средний спин для LCMO25 больше, чем для LCMO30. Следовательно, воп-рос о согласовании низкотемпературных и высокотемпературных магнитных характеристик требует дополнительного исследования.

В предыдущей работе [14] также путем обработки дисперсионных кривых для ферромагнитного соединения LCMO17 были получены значения параметров $J_1$, $J_2$, $J_3$ и $J_4$. Используя результаты статьи [14] и настоящей работы, представим зависимости обменных параметров от концентрации $x$ на рис.3. Обращает на себя внимание немонотонная зависимость $J_1$ и $J_4$, а также большие значения $J_2$ и $J_3$ в LCMO17 по сравнению с LCMO25 и LCMO30. Эти факты также нуждаются в дополнительном обсуждении.

## Список литературы

**Подписи к рисункам**

**Рис.1**. LCMO25. Дисперсионные кривые для основных криисталлографических направлений. Кружки – экспериментальные точки из работы [13], сплошная тонкая линия – расчетная кривая из работы [13], сплошная толстая линия – расчетная кривая настоящей работы.

**Рис.2**. LCMO30. Дисперсионные кривые для основных криисталлографических направлений. Кружки – экспериментальные точки из работы [13], сплошная тонкая линия – расчетная кривая из работы [13], сплошная толстая линия – расчетная кривая настоящей работы.

**Рис.3.** Зависимость обменных параметров от концентрации $x$.

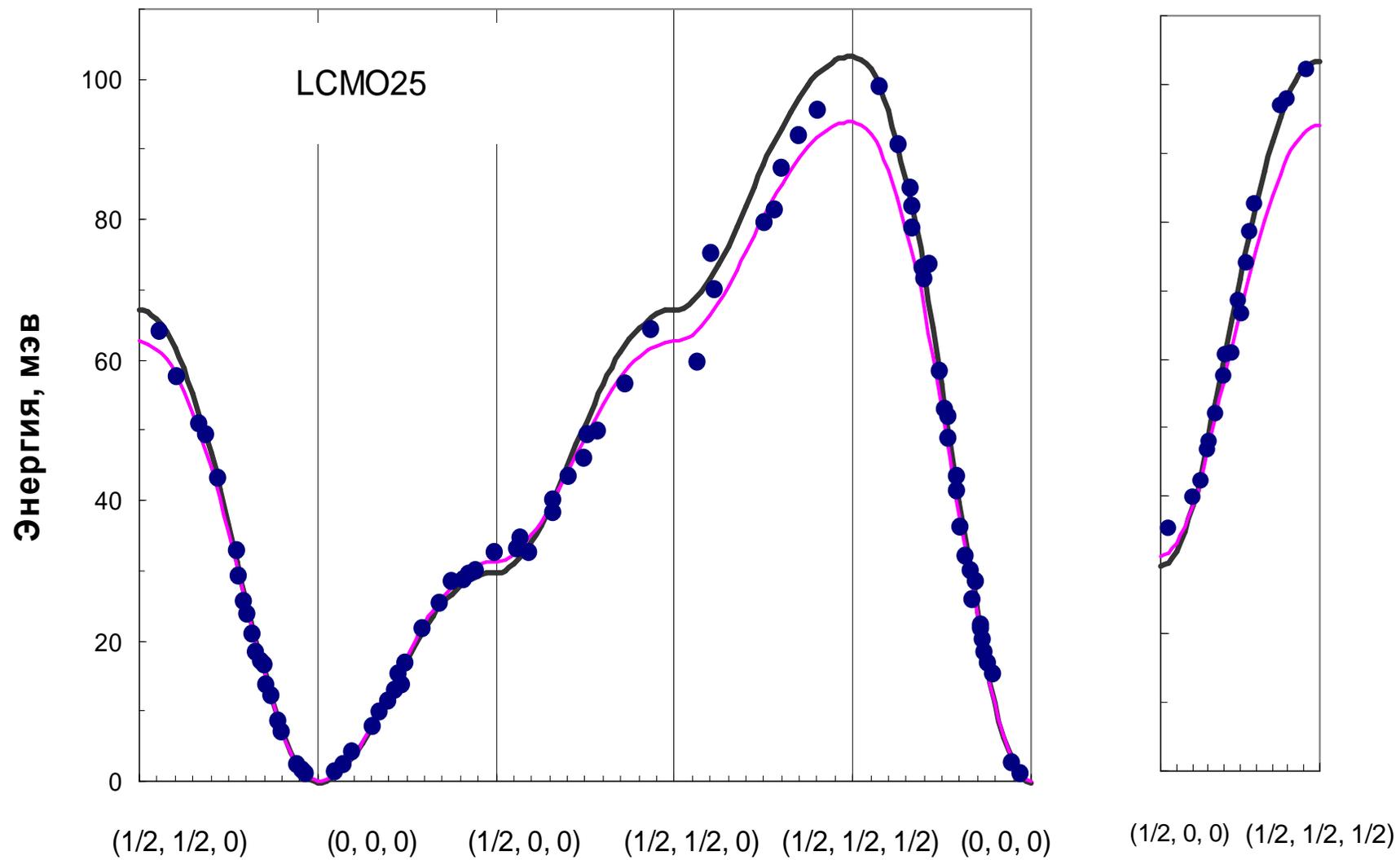

**рис.1**
Б.В.Карпенко, А.В.Кузнецов
Обменные параметры в манганитах лантана

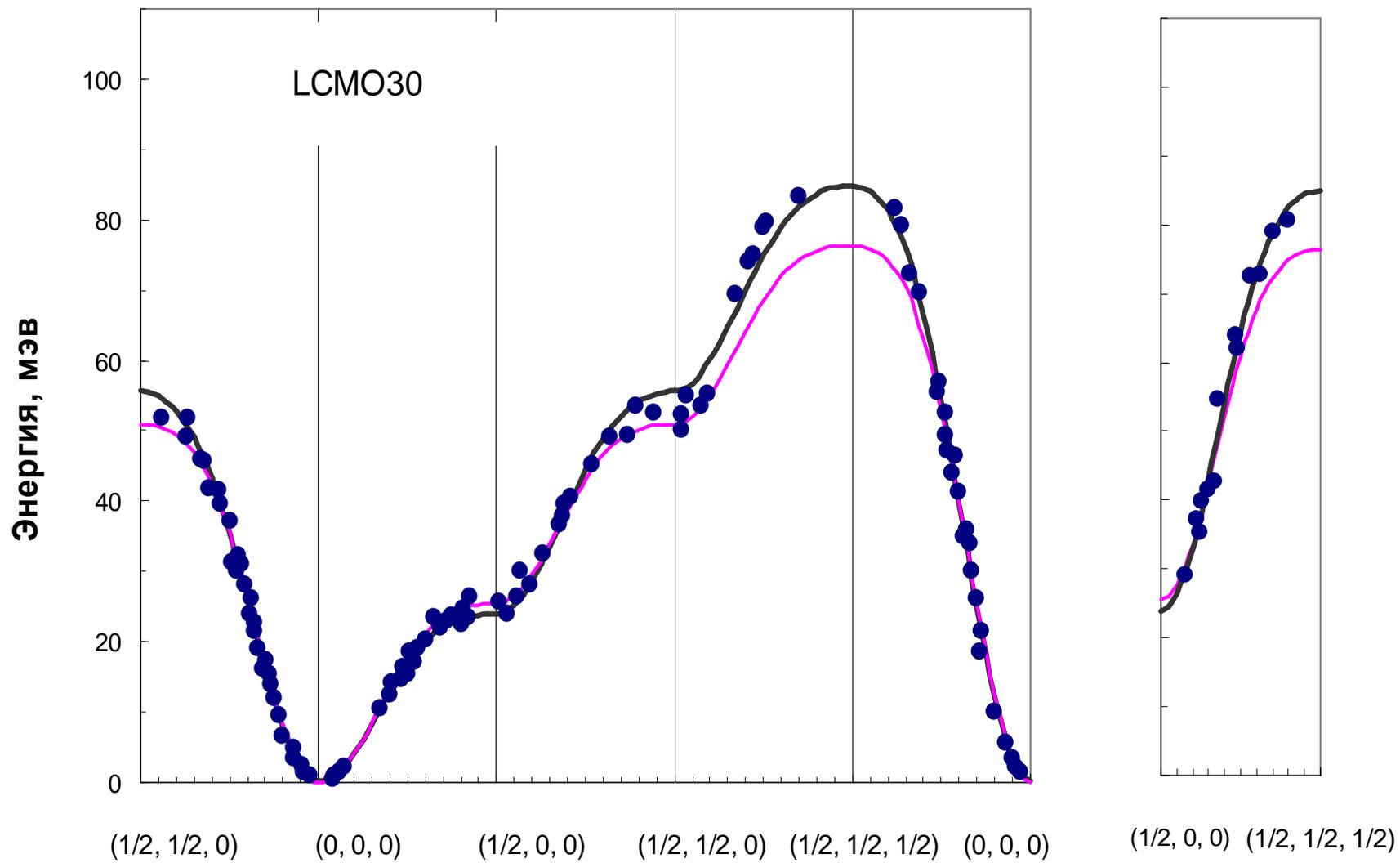

**рис.2**
Б.В.Карпенко, А.В.Кузнецов
Обменные параметры в манганитах лантана

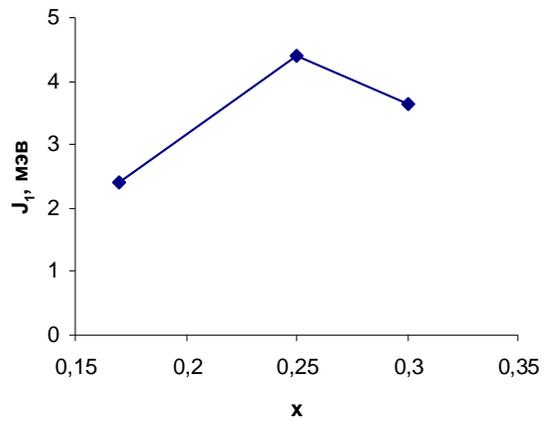
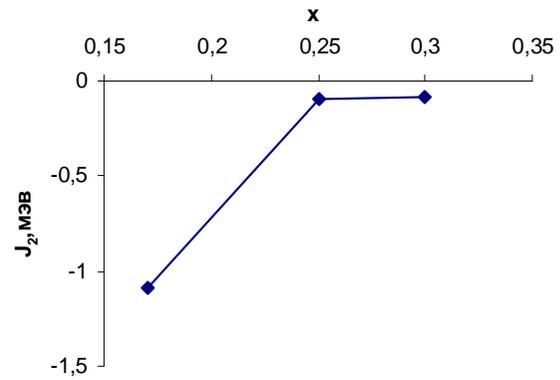
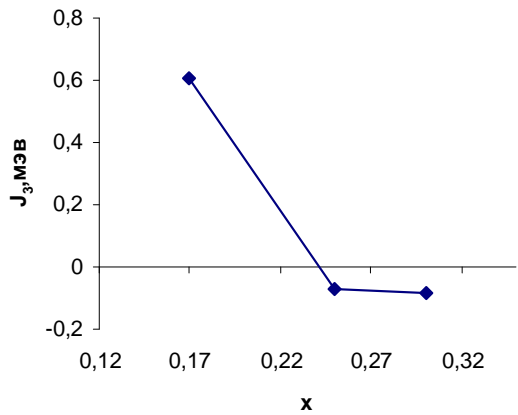
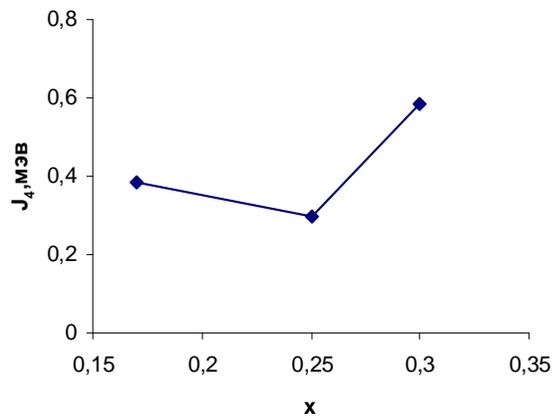

**рис.3**
Б.В.Карпенко, А.В.Кузнецов
Обменные параметры в манганитах лантана